\documentclass[twocolumn]{jpsj3}
\usepackage{times}
\usepackage{color}
\usepackage{amsmath}
\usepackage{graphicx}

\newcommand{\bk}{\mathbf k}

\newcommand{\up}{\uparrow}
\newcommand{\dn}{\downarrow}
\newcommand{\beg}{\begin{equation}}
\newcommand{\en}{\end{equation}}
\newcommand{\dg}{^\dagger}
\newcommand \bea {\begin{eqnarray} }
\newcommand \eea {\end{eqnarray}}

\title{Tunneling in heavy-fermion junctions}

\author{Maxim Dzero\thanks{E-mail address: mdzero@kent.edu} \\
Department of Physics, Kent State University \\
\inst{Kent, OH 44240, U.S.A.}
}
\abst{In this paper I review recent theoretical and experimental advances in understanding of tunneling processes between normal metals and metals containing electrons which occupy partially filled $f$-orbitals. In heavy-fermion materials the effective mass of the quasiparticles far exceeds the bare electron mass  due to strong hybridization between conduction and $f$-orbital states. Kondo lattices form a class of heavy-fermion systems in which an average occupation number of $f$-electron states is close to an integer. Therefore, the tunneling into a Kondo lattice necessarily involves co-tunneling process of a tip electron into an $f$-electron state of a Kondo lattice. This co-tunneling process is manifested in the Fano-lineshape of differential conductance as a function of an applied voltage, which has been routinely observed in recent experiments on various Kondo lattice systems. To illustrate these ideas, I discuss the problem of the tunneling junction when the single particle states in the tip are also a product of hybridization between conduction and $f$-states, i.e. tunneling between two heavy-fermion materials.}

\kword{tunneling, Kondo lattice, correlated electrons}

\begin{document}
\maketitle

\section{Tunneling into a Kondo lattice: overview}
% Brief history
% Co-tunneling of electrons: magnetic impurities in the insulating barrier of the M-I-M junctions
% Tunneling from normal metal -> Kondo lattice
% Tunneling from KL1 -> KL2
% Co-tunneling process in Kondo lattice. No broadening in the f-level ('weakly correlated case')
% Fano-lineshape: review previous theoretical work
% Hamiltonian: PCS vs. STM
In complex materials with atoms containing unfilled $f$-orbitals, interaction between the conduction and $f$-electrons 
leads to the development of a novel electronic states of matter at very low-temperatures. One specific feature of these states is the large effective mass of the electronic excitations. A phenomenologically theory for the emergence of the heavy fermions has been proposed by Coqblin and Blandin \cite{CoqblinBlandin} and Sir Neville Mott \cite{Mott}: strong hybridization between conduction and $f$-electrons produces two bands separated by the hybridization gap, so that renormalized position of the chemical potentials crosses the lower band where the Fermi velocity is significantly reduced implying large effective mass of the electronic excitations. While this picture can also successfully account for the metal-insulator transition in a number of $f$-electron systems with mixed-valence of the $f$-ion - SmB$_6$ and YbB$_{12}$ being the two canonical $f$-orbital semiconductors - the emergence of the coherent band of heavy electrons in both mixed valence and Kondo lattice systems still remained not well understood \cite{PiersReview}. 

Experimentally, one of the main challenges in probing the emergence of the heavy quasiparticles is in the lack of high resolution spectroscopic measurements. Remarkably, this challenge has been overcome in scanning tunneling microscopy measurements \cite{expAynajan1,expSchmidt,expErnst,expAynajan2,JPSmB6} as well as in the point contact spectroscopy \cite{PCSPark,PCSSumiyama,PCSFogel}. Recent tunneling experiments have been convincingly able to trace the formation of the heavy quasiparticles. What is more, momentum and energy resolved tunneling spectra 
visualized not only the formation of the heavy quasiparticles, but also the formation of unconventional superconductivity in 
a prototypical Kondo lattice heavy-fermion superconductor CeCoIn$_5$ \cite{expAynajan1,expAynajan2}. 
Formation of the heavy-particles has also been successfully resolved in more itinerant systems, such as 'hidden order' compound URu$_2$Si$_2$ \cite{expSchmidt} and in a best candidate for correlated topological insulator SmB$_6$ \cite{JPSmB6}.  

Asymmetric or Fano lineshape of the differential tunneling conductance is the basic feature observed in tunneling experiments into Kondo lattice systems. The origin of the Fano lineshape has been explained in a number of 
recent theoretical papers \cite{Marianna,Morr,Balatsky,PCSFogel}. The basic idea for the understanding of the tunneling into a Kondo lattice originates from the earlier models developed by Appelbaum \cite{Appelbaum} and Anderson \cite{Anderson} of the tunneling in metal-insulator-metal (M-I-M) junctions with an insulating layer containing small concentration of magnetic impurities. As they have shown, the tunneling between two metals necessarily involves a process of co-tunneling: an electron from one metal can tunnel directly to another metal, but can also tunnel through a state
on impurity by flipping its spin. Similarly, a tunneling process between the normal metal tip into a Kondo lattice, electron from a tip tunnels into a conduction orbitals as well as into a composite fermion state created by a strong hybridization between conduction and $f$-electrons of a Kondo lattice \cite{Marianna}. 

Interestingly, the observation of the Fano lineshape can actually be used as a fingerprint of strong hybridization between conduction and $f$-electrons even in heavy-fermion systems where Kondo screening competes with an onset of antiferromagnetic order as in CeAuSb$_2$, for example \cite{Hano}. It is important to note, however, that the Fano lineshape of the tunneling conductance appears only for the case when the $f$-electron level acquires a finite lifetime \cite{Balatsky}. Within the currently used mean-field theory approaches \cite{ReadNewns,Piers,Zlatko,NewnsRead,MillisLee} finite lifetime of the $f$-level can either be introduced on the phenomenological level or derived by taking into account the fluctuation corrections to the mean-field theory \cite{MillisLee}. In this paper, I will review these ideas by using the junction between the two heavy fermion metals as an example. I will derive an approximate tunneling Hamiltonian and by resorting to the mean-field approximation I calculate the differential tunneling conductance and discuss the various limiting cases. 

I have organized this paper as follows. In the next Section I will discuss the problem of tunneling between a tip and a host both of which contain states with partially filled $f$-orbitals. In deriving the effective tunneling hamiltonian for that problem 
I will review the main ideas which went into recently developed theories of tunneling into Kondo lattice systems 
\cite{Marianna,Morr,Balatsky}. In Section III I will discuss the open questions as well as possible directions for future research on tunneling involving Kondo lattice materials. 

\section{Tunneling between the two Kondo lattices}
In this Section we discuss the features in differential conductance which would appear in the experiment involving the tunneling contact between the two heavy-fermion metals. At first sight it may seem that as soon as electron leaves 
a heavy-fermion tip it looses all its mass. In what follows I first show that the quasiparticle coherence factors are sufficiently long ranged and a quasiparticle from a tip retains its heavy mass as it reaches the Kondo lattice. Then I proceed with the derivation of the effective tunneling Hamiltonian within the mean-field approximation and obtain analytic expression for the 
differential conductance. At the end of this Section I also discuss the effects of the fluctuations beyond the mean-field theory on the tunneling conductance. 

\subsection{general discussion}
In a heavy-fermion metal single particle states $|i\sigma\rangle=
\hat{p}_{i\sigma}\dg|0\rangle$ in the tip are formed by the superposition of the conduction and localized $f$-states. 
Within the mean-field theory approximation controlled by the parameter $1/N$ where $N$ is given by the degeneracy of the $f$-orbital multiplet, it follows \cite{ReadNewns,Piers,Zlatko,NewnsRead,MillisLee,Hewson,Barzykin}:
\beg
\hat{p}_{i\sigma}=\sum\limits_{l}\left(u_{il}\hat{\tilde f}_{l\sigma}+v_{il}\hat{d}_{l\sigma}\right).
\en
Here $\hat{d}_{l\sigma},\hat{\tilde f}_{l\sigma}$ are an annihilation operators for conduction and $f$-electrons on a site $l$ in the tip and $u_{il},v_{il}$ are heavy-fermion coherence factors:
\beg
\left[\begin{matrix} u({\vec r}_{ij}) \\ v({\vec r}_{ij})\end{matrix}\right]=\sum\limits_{\bk}\left(\begin{matrix} u_\bk \\ v_\bk \end{matrix}\right) e^{-i{\vec k}\cdot{\vec r}_{ij}}, \quad {\vec r}_{ij}={\vec r}_{i}-{\vec r}_j.
\en
Consequently, the momentum dependence of the coherence factors is determined by the spectrum of the conduction electrons $\epsilon_\bk$, $f$-electron single particle energy $\epsilon_f$ and hybridization between them $V_l$. In the simplest tight-binding approximation the spectrum of the conduction electrons is given by $\epsilon_\bk=-2t_c\sum\limits_{i=x,y,z}\cos k_i-\mu$ where $t_c$ is the hopping amplitude and $\mu$ is the chemical potential. The expressions for the  coherence factors are
\beg\label{CohFactors}
\begin{split}
&u_\bk^2=\frac{1}{2}\left(1+\frac{\epsilon_\bk-\epsilon_f}{R_\bk}\right), \quad v_\bk^2=\frac{1}{2}\left(1-\frac{\epsilon_\bk-\epsilon_f}{R_\bk}\right), \\
&R_\bk=\sqrt{(\epsilon_\bk-\epsilon_f)^2+4V_l^2}.
\end{split}
\en

\begin{figure}
\begin{center}
\includegraphics[scale=0.3,angle=0]{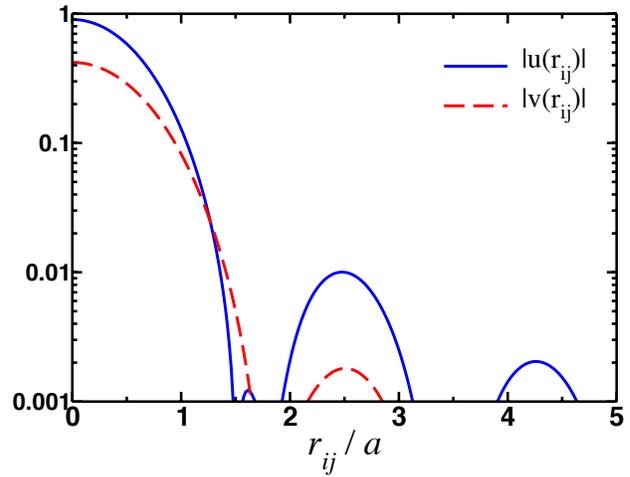}
\end{center}
\caption{(Color online) Spacial dependence of the heavy-fermion coherence factors $u({\vec r}_{ij})$ and $v({\vec r}_{ij})$
as a function of distance between two sites, ${\vec r}_{ij}={\vec r}_i-{\vec r}_j$, in the units of lattice spacing $a$. When the
heavy-electron leaves the tip and the distance between the tip and surface is about several lattice spacings it is clear that
the heavy-electron will retain its composite nature.}
%You can embed figures using the \texttt{\textbackslash includegraphics} command.
%You do not need to separate figures from the main text when you use \LaTeX\ for preparing your manuscript.}
\label{Fig1}
\end{figure}

The heavy-electron, when it tunnels from a tip into a Kondo lattice, will retain its heavy effective mass due to the relatively slow decay of the coherence factors with distance. 
From the analysis of the momentum integrals (\ref{CohFactors}) it is clear that both
$u({\vec r}_{ij})$ and $v({\vec r}_{ij})$ will decay as $\sim 1/r_{ij}^n$ since functions $u_\bk$ and $v_\bk$ are analytic functions of momentum.The values of $\mu, \epsilon_f$ and $V_l$ can be found by employing slave boson mean-field theory \cite{ReadNewns}. In order to compute the spacial dependence of the coherence factors, I have solved the mean-field equations \cite{Barzykin} assuming that $f$-orbital multiplet is sixfold degenerate ($N=6$). The results of the calculation are shown on Fig. \ref{Fig1}. As one can see, the coherence factors extend on the distances of the order of several lattice spacings. Thus, for sufficiently small separation between a tip and a host, there is a finite probability for the composite fermion excitations to tunnel. 

\subsection{approximate tunneling Hamiltonian}
\begin{figure}
\begin{center}
\includegraphics[scale=0.07,angle=0]{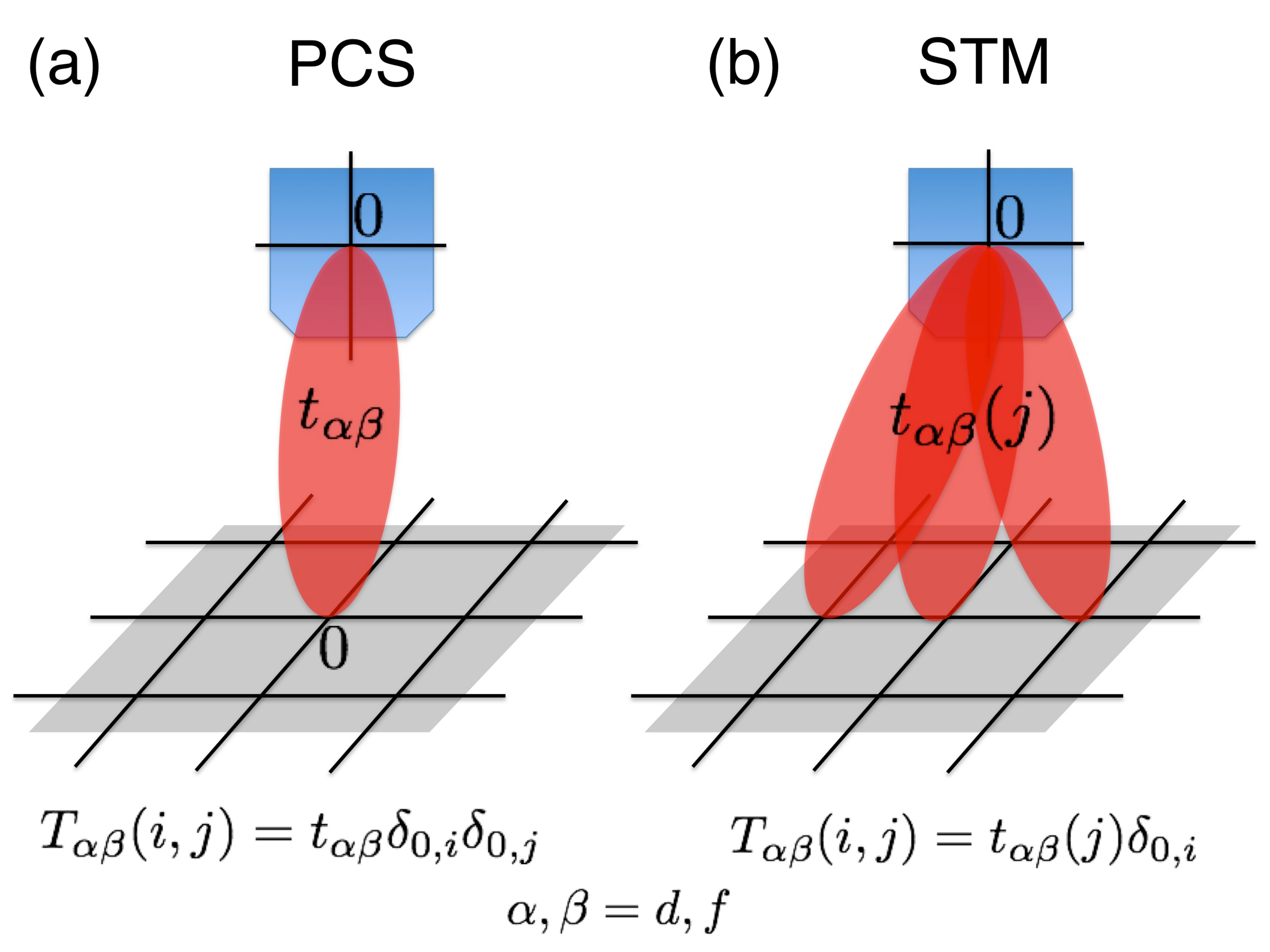}
\end{center}
\caption{(Color online) Two types of tunneling junctions. Panel (a): point contact spectroscopy (PCS) junction. The tunneling matrix elements in this case are diagonal and are non-zero for a single site on a tip and a host. Panel (b): scanning tunneling microscope (STM) junction: tunneling matrix element changes depending on the position of a site in the host, Eq. (\ref{PCS}).}
\label{Fig2}
\end{figure}

To discuss the tunneling between two heavy-fermion metals, I consider the following model Hamiltonian:
\beg\label{Ham}
\begin{split}
\hat{H}=\hat{H}_{AL}^{(1)}+\hat{H}_{AL}^{(2)}+\hat{H}_{t}.
\end{split}
\en
Here $\hat{H}_{AL}^{(a)}$ describe the electrons in the tip ($a=1$) and in the host ($a=2$) correspondingly. We choose them to have the following form of the Anderson lattice model $\hat{H}_{AL}^{(a)}=\hat{H}_{0}^{(a)}+\hat{H}_V^{(a)}$:
\beg\label{HAL}
\begin{split}
\hat{H}_{0}^{(a)}&=\sum\limits_{\bk\sigma}\xi_{a\bk}\hat{d}_{a\bk\sigma}\dg\hat{d}_{a\bk\sigma}+
\sum\limits_{\bk\sigma}\varepsilon_{fa}\hat{f}_{a\bk\sigma}\dg\hat{f}_{a\bk\sigma}\\&+
\frac{U_f}{2}\sum\limits_{i}\hat{f}_{ai\up}\dg\hat{f}_{ai\up}\hat{f}_{ai\dn}\dg\hat{f}_{ai\dn}, \\
\hat{H}_{V}^{(a)}&=V_a\sum\limits_{\bk\sigma}
\left(\hat{d}_{a\bk\sigma}\dg\hat{f}_{a\bk\sigma}+\textrm{h.c.}\right)
\end{split}
\en
where the first term in $\hat{H}_0$ describes conduction electrons and the remaining two terms describe the $f$-electrons,
while $\hat{H}_V$ accounts for the hybridization between conduction and $f$-electrons. Few comments are in order. To simplify our subsequent discussion here I consider the Kramers doublets for the ground state of the $f$-electrons and label them the same way as the spin state of the conduction electrons, $\sigma=\up,\dn$. In (\ref{HAL}) we have also ignored that fact that conduction electrons orbitals usually have $l=0,1,2$ orbital number, which makes the hybridization matrix element with $f$-electrons ($l=3$) nonlocal. Lastly, the third term in (\ref{HAL}) accounts for the tunneling events. In accord with our general discussion above, the non-local form of the
coherence factors allows for the tunneling events not only between the conduction orbitals, but also between the $f$-orbitals.
Hence, the most general form of the tunneling Hamiltonian is:
\beg
\begin{split}
\hat{H}_{t}&=\sum\limits_{ij}\left(\hat{d}_{1i\sigma}\dg T_{dd}(i,j) \hat{d}_{2j\sigma}+
\textrm{h.c.}\right)\\&+
\sum\limits_{ij}\left(\hat{\psi}_{1i}\dg\left[
\begin{matrix} 0 & T_{df}(i,j) \\ 
T_{fd}(i,j) & T_{ff}(i,j) \end{matrix}\right]\hat{\psi}_{2j}+
\textrm{h.c.}\right)\equiv \\&\equiv \hat{H}_{tun}+\hat{H}_{co-tun},
\end{split}
\en
where $\hat{\psi}_{ai}\dg=\left(\hat{d}_{aj}\dg, ~\hat{f}_{aj}\dg\right)$ and we have omitted the spin index for brevity. We also need to keep in mind that the tunneling amplitude $T_{ff}$ between the $f$-orbitals is much smaller than the rest of the tunneling matrix elements and will be ignored. In what follows we consider the simplest case,  when the tunneling matrix elements are diagonal in site indices \cite{TunNice},Fig. \ref{Fig2}:
\beg\label{PCS}
T_{\alpha\beta}(i,j)=T_{\alpha\beta}\delta_{i,0}\delta_{j,0}, \quad \alpha,\beta=d,f.
\en
%%%%% Start of Fig 2 %%%%%
%%%%% End of Fig 2 %%%%%

For the case of a junction between two Kondo lattices - the main subject of this Section - the Hubbard repulsion between the $f$-electrons (\ref{HAL}) is the largest energy scale of the problem. When both $|\varepsilon_{fa}|$ and 
$\varepsilon_{fa}+U_f$ are much larger than the conduction electrons density of states $\rho_F$ times the square of the hybridization amplitude, $|\varepsilon_{fa}|,\varepsilon_{fa}+U_f\gg \textrm{max}[\rho_F|V|^2,\rho_F|t_{\alpha\beta}|^2]$, the doubly occupied states on $f$-sites can be integrated out by means of the Schrieffer and Wolff transformation \cite{SchriefferWolff}. Specifically, the effective tunneling Hamiltonian can be obtained by the unitary transformation
$\hat{\tilde H}=e^{\hat{\cal S}}\hat{H}e^{-\hat{\cal S}}=\hat{H}_0+\hat{H}_{eff}$
and the anti-hermitian operator $\hat{\cal S}$ must be determined from
\beg\label{S}
\left[\hat{\cal S},\hat{H}_0^{(1)}+\hat{H}_0^{(2)}\right]=
-\hat{H}_V^{(1)}-\hat{H}_V^{(2)}-\hat{H}_{co-tun}.
\en
We note, that $\hat{\cal S}$, Eq. (\ref{S}),  will depend on hybridization amplitudes in the tip and the host Anderson lattices as well as the tunneling amplitudes between the conduction electron in the tip and an $f$-electron in the host and visa versa. 
As a result of this transformation, we obtain an effective Hamiltonian $\hat{H}_{eff}$ by retaining terms up to the second order in $V$ and/or $t_{\alpha\beta}$. Naturally, $\hat{H}_{eff}$ will be given by the sum of the Kondo lattice Hamiltonians for both the tip and the host electrons:
\beg\label{HKL}
\begin{split}
\hat{H}_{KL}^{(a)}&\approx \sum\limits_{\bk\sigma}\xi_{a\bk}\hat{d}_{a\bk\sigma}\dg\hat{d}_{a\bk\sigma}\\&+J_{K}^{(a)}\sum\limits_{i;\alpha\beta}\hat{d}_{ai\alpha}\dg
\left({\vec S}_{ai}\cdot{\vec \sigma}_{\alpha\beta}\right)\hat{d}_{ai\beta}
\end{split}
\en
and the effective tunneling Hamiltonian which we write down as a sum of the two terms $\hat{H}_{tun}=\hat{H}_{tun}^{(d)}+\hat{H}_{tun}^{(f)}$, where:
\beg\label{Htund}
\hat{H}_{tun}^{(d)}=\sum\limits_{i\sigma}\left\{t_{d}\hat{d}_{1i\sigma}\dg\hat{d}_{2i\sigma}+
t_{d}^*\hat{d}_{2i\sigma}\dg\hat{d}_{1i\sigma}\right\}\delta_{i,0}
\en
with $t_d=T_{dd}$ describes the tunneling between the conduction states in the tip and the conduction states in the host. 
Consequently, the second term in the tunneling Hamiltonian
\beg\label{Htunf}
\begin{split}
{H}_{tun}^{(f)}=&\left[J_{12}\sum\limits_{i\sigma\alpha}
\overbrace{\hat{d}_{1i\sigma}\dg\left({\vec S}_{1i}\cdot{\vec \sigma}_{\sigma\alpha}\right)\hat{d}_{2i\alpha}}^{tip \rightarrow host ~~co-tunneling}\right.\\&\left.+J_{21}\overbrace{\hat{d}_{1i\sigma}\dg\left({\vec S}_{2i}\cdot{\vec \sigma}_{\sigma\alpha}\right)\hat{d}_{2i\alpha}}^{host\rightarrow tip ~~co-tunneling}\right]\delta_{i,0}
+\textrm{h.c.},
\end{split}
\en
where ${\vec S}_{ai}=\frac{1}{2}\hat{f}_{ai\alpha}\dg{\vec \sigma}_{\alpha\beta}\hat{f}_{ai\beta}$ are local moments in a
tip and a host, $J_{12}$ and $J_{21}$ are corresponding exchange coupling constants proportional to the tunneling matrix
elements $t_{df}$. A crucial difference with the models considered earlier \cite{Marianna,Morr,Balatsky} is the presence
of the co-tunneling terms proportional to $J_{21}$, which account for the tunneling of the composite electrons in the tip into conduction electron orbitals of the host. Before we proceed with the calculation of the tunneling current, we note that in deriving an effective
Hamiltonian we have ignored the tunneling events between the predominantly localized $f$-electrons as well as other terms generated by the Schrieffer-Wolf transformation, which in principle could affect the tunneling current. However, it is known
that these terms can be safely ignored in the problem of the tunneling from normal metal into a Kondo lattice \cite{Marianna} 
since the model Hamiltonian $\hat{H}_{eff}$ with $J_{21}=0$, Eqs. (\ref{Htund},\ref{Htunf}), 
provides more than adequate description of the available experimental data 
\cite{expAynajan1,expSchmidt,expErnst,expAynajan2,JPSmB6}. At the same time, for the analysis of tunneling experiments into a superconducting Kondo lattice \cite{expAynajan2} these terms may actually be important, especially for probing unconventional Cooper pairing mechanisms. 

\subsection{tunneling current}
The tunneling current is defined by the rate of change in the number of conduction electrons in a tip, $I(V)=|e|\langle\dot{\hat N}_{tip}\rangle$ with $\hat{N}_{tip}=\sum\limits_{i\sigma}\hat{d}_{1i\sigma}\dg\hat{d}_{1i\sigma}$ and averaging is performed in the grand canonical ensemble with full Hamiltonian $\hat{H}_{eff}$. Clearly, the nonzero value for the current is furnished by the presence of the tunneling terms (\ref{Htund},\ref{Htunf}) in the Hamiltonian. Moreover, the substantial progress can be made by adopting the large-$N$ limit approximation for the Kondo lattice \cite{PiersReview}. Within this approximation, the composite fermion operators entering into the expression for tunneling current can be expressed as a single fermionic operator according to \cite{Marianna}:
\beg\label{composite}
\begin{split}
&\sum\limits_{\beta}\left({\vec S}_{1i}\cdot{\vec \sigma}_{\alpha\beta}\right)\hat{d}_{1i\beta}\to
\frac{\tilde{t}_{f}}{J_{21}}\hat{f}_{1i\alpha}, \\
&\sum\limits_{\beta}\left({\vec S}_{2i}\cdot{\vec \sigma}_{\alpha\beta}\right)\hat{d}_{2i\beta}\to
\frac{t_{f}}{J_{12}}\hat{f}_{2i\alpha}.
\end{split}
\en
In what follows, without loss of generality we take $t_f\approx\tilde{t}_f$. 
The rest of the calculation employs the standard methods \cite{TunNice} and we will not provide the details here. The resulting expression for the current reads:
\beg\label{TunCurrent}
\begin{split}
I(V)&=I_{tun}(V)+\delta I(V), \\
I_{tun}(V)=
\frac{2\pi e}{\hbar}&\int\limits_{-\infty}^{\infty}d\omega[n_F(\omega-eV)-n_F(\omega)]\\&\times
\sum\limits_{\alpha=1,2}\Pi_{\alpha,\beta\not=\alpha}(\omega-eV,\omega), \\
\delta I(V)=\frac{2\pi e}{\hbar}&\int\limits_{-\infty}^{\infty}d\omega[n_F(\omega-eV)-n_F(\omega)]\\&\times
\delta\Pi(\omega-eV,\omega),
\end{split}
\en
Here functions $\Pi_{\alpha,\beta}$ and $\delta\Pi$ are defined as follows:
\beg
\begin{split}
&\Pi_{\alpha,\beta}(\omega,\omega')=\rho_{\alpha c}(\omega)\left[t_c^2\rho_{\beta c}(\omega')+2t_ft_c\rho_{\beta m}(\omega')\right.\\&\left.+t_f^2\rho_{\beta f}(\omega')\right], \\
&\delta\Pi(\omega,\omega')=2t_f^2\rho_{1m}(\omega)\rho_{2m}(\omega')-t_c^2\rho_{1c}(\omega)\rho_{2c}(\omega').
\end{split}
\en
with $\rho_{\alpha a}(\omega)$, ($\alpha=1,2;a=c,f,m$) being determined by the single particle propagators 
of the corresponding Kondo lattices. Within the mean-field approximation we have:
\beg\label{rhos}
\begin{split}
&\rho_{ac}(\omega)=-\frac{1}{\pi}\textrm{Im}\sum\limits_{\bk}\frac{1}{\omega^+-\xi_{\alpha\bk}-\frac{|V_\alpha|^2}{\omega^+-\lambda_\alpha}}, \\ &\rho_{af}(\omega)=-\frac{1}{\pi}\textrm{Im}\sum\limits_{\bk}\frac{1}{\omega^+-\lambda_\alpha-\frac{|V_\alpha|^2}{\omega^+-\xi_{\alpha\bk}}}, \\
&\rho_{am}(\omega)=-\frac{1}{\pi}\textrm{Im}\sum\limits_{\bk}\frac{V_\alpha/(\omega^+-\lambda_\alpha)}{\omega^+-\xi_{\alpha\bk}-\frac{|V_\alpha|^2}{\omega^+-\lambda_\alpha}}
\end{split}
\en
where $\lambda_\alpha$ denotes renormalized position of the $f$-level and we have assumed that $\omega^+=\omega+i\delta$ in Eqs. (\ref{rhos}) is a complex number with an infinitesimally small and positive imaginary part. 

\begin{figure}
\begin{center}
\includegraphics[scale=0.3,angle=0]{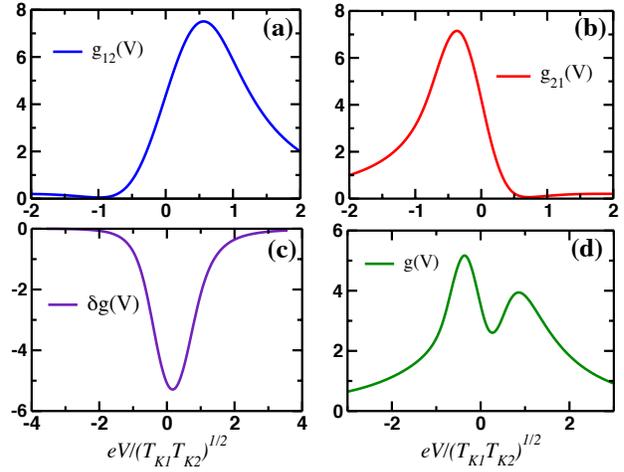}
\end{center}
\caption{(Color online) Three contributions to the differential tunneling conductance 
$g(V)=dI/dV=g_{12}(V)+g_{21}(V)+\delta g(V)$, where $I(V)$ is given by Eq. (\ref{TunCurrent}), together with $g(V)$ (in arbitrary units) are shown. Here $g_{12}$ is determined by the $a=1$ term in the expression for $I_{tun}(V)$ and is governed by the tunneling and cotunneling processes between conduction orbitals of the tip and conduction and $f$-electron orbitals in the host. Consequenctly, $g_{21}$ is given by the $a=2$ contribution to $I_{tun}(V)$ and describes the tunneling and cotunneling events between the conduction orbitals in the host and the conduction and $f$-electron orbitals in the tip. Lastly, $\delta g(V)=d\delta I/dV$ is an interference term between the co-tunneling events. All contributions to the differential tunneling conductance are plotted as a function of voltage in the units of the $\sqrt{T_{K1}T_{K2}}$, where $T_{K1,2}$ are the corresponding Kondo lattice coherence temperatures of the tip and the host.}
\label{Fig3}
\end{figure}

The first term in the expression for the current (\ref{TunCurrent}) has a simple physical interpretation: it describes the tunneling events between the conduction orbitals of the tip (host) into the conduction and composite fermion states of the host (tip). When one neglects the finite width of the $f$-electron level, the local density of states in the tip and the
host has two peaks as a function of frequency \cite{Marianna,Balatsky}. If we now include the corrections due to 
the fluctuations of the mean-field amplitude \cite{MillisLee}, 
which are proportional to $1/N$, where $N$ is the degeneracy of the $f$-level, 
the $f$-electron enegy acquires an imaginary part which depends both on momentum and frequency. 
To simplify our discussion, one can include the constant imaginary part, i.e. replace $\lambda_a\to\lambda_a-i\gamma$, where $\gamma$ is of the order of Kondo lattice coherence temperature. Consequently, each of the two terms (second equation in (\ref{TunCurrent})) contributing to the differential tunneling conductance 
$g_{tun}(V)=dI_{tun}/dV$ has an asymmetric shape as a function of voltage due to the co-tunneling processes between the conduction and composite fermion states \cite{Marianna,Morr,Balatsky}. Therefore, we can approximately write
\beg
\begin{split}
g_{tun}(\varepsilon)&\approx\rho_{F1}\frac{\left(q_1\Gamma_1-\varepsilon-\epsilon_{1}\right)^2}{\left(\varepsilon+\epsilon_1\right)^2+\Gamma_1^2}\\&+
\rho_{F2}\frac{(q_2\Gamma_2+\varepsilon-\epsilon_{2})^2}{(\epsilon-\epsilon_2)^2+\Gamma_2^2}.
\end{split}
\en
Here $q_a$, $\epsilon_a$ and $\Gamma_a\sim 2T_K^{(a)}$ are corresponding parameters, which determine the Fano lineshape, while $\rho_{Fa}$ denotes the conduction band density of states in the tip ($a=1$) and the host ($a=2$). 
Moreover, the direct calculation shows that the second contribution $\delta g(V)=d\delta I/dV$ to the tunneling conductance $dI/dV$ (\ref{TunCurrent}) remains slightly asymmetric and does not have a characteristic Fano lineshape, see Fig. \ref{Fig3}(c).
Thus, we see that for the Kondo lattice materials with comparable parameters, such as hybridization, $f$-level position and
the coherence temperature, the tunneling conductance is drastically different than the one for the tunneling involving normal and heavy-fermion metals. Specifically, for the tunneling contact involving identical Kondo lattice systems, the tunneling conductance will have a symmetric form around zero bias. 
%%%%% Start of Fig 3 %%%%%
%%%%% End of Fig 3 %%%%%

\section{Conclusions}
In this paper, I have reviewed some of the recent theoretical advances in the problem of tunneling between normal and
heavy-fermion or Kondo lattice systems. The main feature in the tunneling conductance between normal and Kondo lattice
is the presence of the asymmetry well described by the Fano lineshape. The origin of the asymmetry lies in the tunneling
processes of the uncorrelated electrons in the tip into the composite fermionic states created by the strong hybridization between the conduction and $f$-electron states in the Kondo lattice. In a junction between the two Kondo lattice the asymmetric features in the tunneling conductance are greatly suppressed compared to the normal metal-Kondo lattice junctions. 
\subsection{Acknowledgements} 
I gratefully acknowledge very useful discussions with Wan Kyu Park, Laura Greene and Piers Coleman. 
This work was financially supported by Ohio Board of Regents (grant OBR-RIP-220573) at KSU, 
the U.S. National Science Foundation I2CAM International Materials
Institute Award, Grant DMR-0844115 and the ICAM Senior Fellowship Grant.

\end{document}